# New signatures on dissipation from fission induced by relativistic heavy-ion collisions[a]

B. Jurado[1b], C. Schmitt[1], K.-H. Schmidt[1], J. Benlliure[2], T. Enqvist[1c], A. R. Junghans[3], A. Kelić[1], F. Rejmund[1d],

[1)]*GSI, Planckstr.1, 64291 Darmstadt, Germany*
[2)]*Universidad de Santiago de Compostela, 15706 Santiago de Compostela, Spain*
[3]*Forschungszentrum Rossendorf, Postfach 510119, 01314 Dresden, Germany*

**Abstract:**
Fissile nuclei with small shape distortion relative to the ground-state deformation and with low angular momentum were produced in peripheral heavy-ion collisions. Under the conditions of small shape distortions and low angular momentum, the theoretical description of the fission process can be considerably simplified, and the relevant information on dissipation can be better extracted than in conventional experiments based on fusion-fission reactions. In addition, this experimental approach induces very high excitation energies, a condition necessary to observe transient effects. The experimental data were taken at GSI using a set-up especially conceived for fission studies in inverse kinematics. This set-up allowed determining three observables whose sensitivity to dissipation was investigated for the first time: the total fission cross sections of $^{238}$U at 1 $A$ GeV as a function of the target mass, and, for the reaction of $^{238}$U at 1 $A$ GeV on a $(CH_2)_n$ target, the partial fission cross sections and the partial charge distributions of the fission fragments. The comparison of the new experimental data with a reaction code adapted to the conditions of the reactions investigated leads to clear conclusions on the strength of dissipation at small deformation where the existing results are rather contradictory.

## 1. Introduction

Dissipation is a fundamental process in nuclei that determines the time an excited nucleus needs to populate the available deformation space and to reach equilibrium. The concept of dissipation was already introduced by Kramers [1] more than sixty years ago, but the success of the transition-state model of Bohr and Wheeler [2] prevented his idea to establish. However, it was found by different groups in the 80's [3, 4] that measured pre-scission neutron multiplicities were much larger than the predictions of the transition-state model. This discrepancy was interpreted as an indication that the deexcitation process of a highly excited heavy nucleus is a dynamical process. Recent efforts have been made to develop fundamental models of fission based

---

exclusively on the individual interaction between the nucleons [5]. Unfortunately, a complete dynamical description of this deexcitation process in terms of a purely microscopic theory is not possible to the present day due to the large number of degrees of freedom involved. For this reason, most of the current theoretical models are transport theories [6] that try to portray the process using a small number of variables. In these theories one distinguishes between collective (or macroscopic) and intrinsic (or microscopic) degrees of freedom. The latter are not considered in detail but in some average sense as a heat bath. The collective degrees of freedom of the nucleus correspond to the coordinate motion of part or all the nucleons, e.g. vibrations and rotations. The intrinsic degrees of freedom are the individual states of the nucleons. The process of energy transfer between the collective degrees of freedom and the heat bath is denominated dissipation. It is quantified by the reduced dissipation coefficient $\beta$, which is defined by the equation:

$$\frac{dE_{coll}}{dt} = \beta \left[ E_{coll}^{eq} - E_{coll} \right] \quad (1)$$

where $E_{coll}$ and $E_{coll}^{eq}$ are the average excitation energy in the collective degree of freedom at a given time $t$ and at thermal equilibrium, respectively. From equation (1) it follows that the reduced dissipation coefficient measures the relative rate with which the excitation energy of the collective degree of freedom changes. Thus, $\beta$ rules the relaxation of the collective degrees of freedom towards thermal equilibrium.

The fission process represents the clearest example of a large-scale collective motion. In addition, the two fission fragments that result from the excitation of this collective motion represent a clear signature that allows identifying the mechanism univocally. In 1940 Kramers [1] suggested describing fission as a dissipation process where the evolution of one or more collective degrees of freedom is given by the interaction with the surrounding medium formed by the individual nucleons. Such process can be described by the Fokker-Planck equation (hereafter FPE) [7], where the reduced dissipation coefficient $\beta$ is a parameter. Kramers found that the stationary solution of the FPE leads to a reduction of the fission width compared to the transition-state model prediction of Bohr and Wheeler [2] by a factor

$$K = \sqrt{1+\gamma^2} - \gamma \quad (2)$$

with

$$\gamma = \frac{\beta}{2\omega_{sad}} \quad (3)$$

where $\omega_{sad}$ is the frequency of the harmonic-oscillator potential that osculates the fission barrier at the saddle point.

More than forty years later, Grangé, Jun-Qing and Weidenmüller [8] and others investigated theoretically the influence of nuclear dissipation on the fission time scale for an excited, initially undeformed, system. From the numerical solution of the time-dependent FPE they derived that the fission-decay width $\Gamma_f(t)$ needs some time to set in and to finally reach the stationary value $\Gamma_K$ predicted by Kramers. In reference [9] the time evolution of the fission width was characterised by the transient time $\tau_{tran}$, defined as the time in which $\Gamma_f(t)$ reaches 90% of its asymptotic value.

The relation between the transient time and the reduced dissipation coefficient $\beta$ can be extracted from the numerical solution of FPE. Reference [9] estimates it by the following analytical approximation:

$$\tau_{tran} = \frac{1}{\beta} ln\left(\frac{10B_f}{T}\right) \quad \text{for } \beta < 2\omega_g$$

$$\tau_{tran} = \frac{\beta}{2\omega_g^2} ln\left(\frac{10B_f}{T}\right) \quad \text{for } \beta > 2\omega_g$$

(4)

where $B_f$ is the fission barrier, $T$ is the nuclear temperature and $\omega_g$ is the oscillator frequency at the ground state.

Intense work has been performed during the last three decades to understand the origin of dissipation in nuclei as well as its variation with deformation and temperature. Nonetheless, the current theories on dissipation give rather contradictory results [10]. A similar controversial situation is found on the experimental side. Several works [11, 12] have already remarked the difficulty of finding experimental signatures sensitive to the temperature dependence of dissipation. With regard to the deformation dependence, for the large-deformation regime clear signatures of dissipation have been found [12, 13, 14], whereas the situation is rather uncertain for the regime of smaller deformations inside the saddle point. While some studies find clear effects with $\beta = 4 \cdot 10^{21} s^{-1}$ [14] and $\beta = 6 \cdot 10^{21} s^{-1}$ [12] or with a fission delay time of $10^{-19}$ to $10^{-20}$ s [15], other works point to weak effects with $\tau_{tran} \leq 10 \cdot 10^{-21}$s [16], $\tau_{tran} < 1 \cdot 10^{-21}$s [17], $\beta = 2 \cdot 10^{21} s^{-1}$ [18] or to no dissipation effects at all [19, 20]. To clarify the situation, new observables, sensitive to dissipation in the small-deformation regime, should be introduced, and alternative mechanisms to induce fission are needed.

The majority of the experimental approaches dedicated to the study of dissipation are based on nucleus-nucleus collisions at energies that range from 5 $A$ MeV to about 100 $A$ MeV. Among the experimental observables studied in this type of reactions the most common are the particle [21] and $\gamma$-ray [22] multiplicities, the angular, mass and charge distributions of the fission fragments [23], and the fission and evaporation-residue cross sections. Except for the fission and evaporation-residue cross sections, all these observables give information on dissipation on the whole path from the ground-state deformation to scission, but they do not allow exploring the deformation range from the ground state to the saddle point independently. In addition, fusion-fission and quasi-fission reactions, which are mostly used, induce initial composite systems with large deformation, and therefore they do not offer optimum conditions for extracting the relevant information at small deformation. In these cases, a reliable interpretation of the experimental signatures involves using elaborated dynamical codes that describe the complete evolution of the composite system along the dynamical trajectory. Moreover, such codes have to deal with the large angular momenta of the composite system and to include the effects of fluctuations around the mean trajectory. Contrary to fusion-fission and quasifission reactions, antiproton annihilation [24, 25, 26, 17], very peripheral transfer reactions [27] and spallation reactions [18] lead to fissioning nuclei with small deformation and small angular momentum, simplifying the theoretical description considerably. High excitation energies are necessary to observe transient effects most clearly. Indeed, since the transient time is expected to be small, the excitation energy of the nucleus should be high enough for the statistical decay time to be comparable with the transient time. Thus, an experimental approach is required that produces highly-excited heavy nuclei with large cross sections. The observables most commonly investigated using these reaction mechanisms are total fission and evaporation-residue cross sections. However, we have shown in references [28, 29] that these observables are not sufficient to answer questions like the temperature dependence of dissipation or the influence of an approximate description of the time dependence of the fission decay width on the magnitude of the dissipation strength. In the present work, we will discuss an experimental method based on peripheral heavy-ion collisions at relativistic energies that offers ideal conditions for investigating dissipation at small deformation.

## 2. Peripheral heavy-ion collisions at relativistic energies

In the ideal scenario for our investigation, the heavy nucleus produced is highly excited with only little shape distortion. In addition, the angular momentum induced should be small in order to avoid additional complex influence on the fission process. These specific initial conditions can be achieved by applying a projectile-fragmentation reaction, i.e. a very peripheral nuclear collision in inverse kinematics with relativistic heavy ions. The description of such reactions can be divided into three main stages: First, the collision takes place leading to a prefragment, then thermal equilibrium in the intrinsic degrees of freedom is established and, finally, the resulting equilibrated system decays in a competition of particle emission and fission. A scheme of these three stages is shown in figure 1.

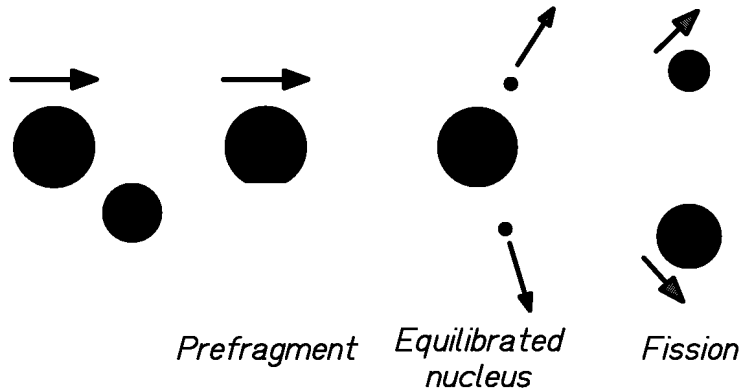

**Figure 1:** Scheme of the different stages of a very peripheral heavy-ion collision at relativistic energies that leads to fission.

The characteristics of the prefragment are described by the abrasion model [30, 31, 32], which is documented in section 5.2. Nuclear collisions at bombarding energies well above the Fermi energy can be considered as quasi-free nucleon-nucleon collisions. Hence, a peripheral collision of the relativistic heavy projectile with the target essentially removes a number of nucleons from the projectile and the target nucleus. The shape of the prefragment is almost not distorted, the root-mean-square value of the angular-momentum distribution of the prefragment varies from 10 to 20 $\hbar$ (as shown by numerical calculations [33]), and its mean excitation energy is given by the number of nucleons abraded. It has been found experimentally [32] that on the average 27 MeV excitation energy per nucleon abraded is induced. As mentioned in the introduction, similar initial conditions can be reached by relativistic proton-nucleus collisions [18] and by the annihilation of antiprotons [34, 24] at the nuclear surface. However, a comparison between model calculations based on the intranuclear cascade [35] and the abrasion model [31, 32] shows that, if the same fissioning nucleus is produced, the angular momentum induced by proton reactions is approximately three times larger than the one induced by peripheral fragmentation reactions with heavy nuclei. In addition, peripheral fragmentation reactions populate higher excitation energies more strongly than proton or antiproton-induced reactions do. This is shown in figure 2 for the case of antiproton reactions, where the experimental differential total reaction cross section $d\sigma/dE^*$ for the system $\bar{p}$ (1.2 GeV) + $^{238}$U taken from reference [34] is compared with a calculation according to the abrasion model for the fragmentation reaction of $^{238}$U (1 $A$ GeV) + Pb. Recent experimental results [36] indicate that at temperatures exceeding values around 5.5 MeV thermal instabilities set in, eventually leading to multifragmentation. This means that the system undergoes a simultaneous break-up. This phenomenon is considered as well in the theoretical model used to interpret these reactions, as discussed below. As can be seen in figure 2, while the reaction cross section induced by antiproton annihilation starts to decrease steeply at excitation energies of about 500 MeV, this range of energies remains strongly populated by fragmentation reactions induced by heavy nuclei. This is of great advantage not only because it enables observing the effects of the dynamical delay induced by dissipation and analysing a possible temperature dependence, but also because it allows investigating the interplay between dissipation and thermal instabilities in the inhibition of fission at high excitation energies, see reference [29].

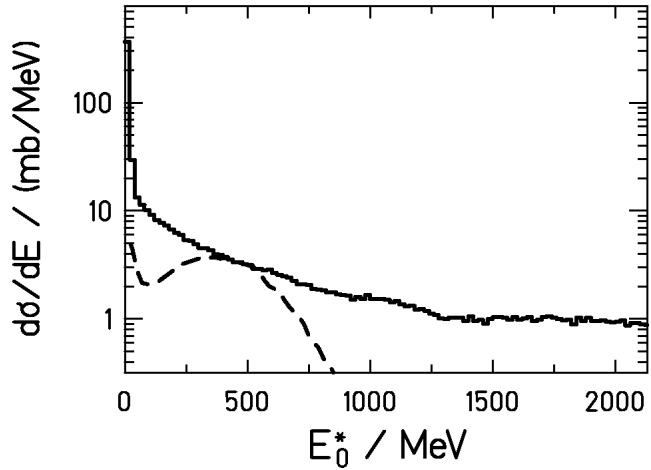

**Figure 2:** Total reaction cross section as a function of the excitation energy induced right after the collision, before an eventual pre-equilibrium or break-up process and the consecutive sequential decay. The dashed line corresponds to the reaction $\bar{p}$ (1.2 GeV) + $^{238}$U measured in reference [34], and the full line is a calculation performed with the abrasion model [31, 32] for the reaction $^{238}$U (1 $A$ GeV) + Pb.

## 3. Experimental set-up

To investigate the reactions described above, the GSI facilities were used. The heavy-ion synchrotron SIS delivered an intense relativistic $^{238}$U beam of 1 $A$ GeV to the experimental set-up, schematically illustrated in figure 3. This set-up, especially designed for fission studies in inverse kinematics, consisted of a scintillation detector [37], two MUltiple-Sampling Ionisation Chambers (MUSICs) [38], a double ionisation chamber and a time-of-flight wall. The first scintillation detector supplied the start signal for the time-of-flight measurement and served as a trigger. The target was located in between the two MUSICs. This configuration worked as an active target and provided the information to distinguish reactions in the target material from reactions in other layers of matter. The double IC recorded the energy-loss signals of both fission fragments separately, and the time-of-flight wall provided the stop signal for their time-of-flight measurement. In addition its granularity of 31 horizontal zones due to the partial overlap of 15 paddles allowed applying a multiplicity filter on the events, which should be two for fission fragments. This set-up was a modified version of a previous one [39] that was conceived to study electromagnetic-induced fission in a lead target. The even-odd structure of the charge yields of the fission fragments measured there [40] was used to study dissipation in cold nuclei [41]. In the present case, we wanted to focus on dissipation at higher excitation energies. Therefore, we optimised the set-up for investigating nuclear-induced fission in different targets. In the previous experiment [39], a subdivided scintillation detector, mounted in front of the double IC, served as a fast trigger to reduce the load of the data acquisition. In the present measurement, this detector was removed to avoid corrections due to secondary reactions of the fission fragments when passing through it.

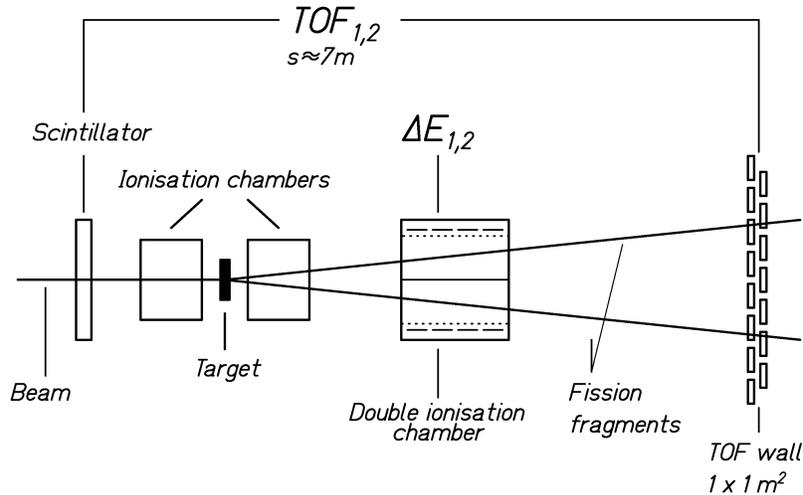

**Figure 3**: Experimental set-up for fission studies in inverse kinematics.

The double ionisation chamber (IC) was conceived according to the kinematics of fission residues in inverse kinematics. It consisted of two independent counting-gas volumes, separated horizontally by a common cathode. The set-up provided a detection efficiency for fission products of approximately 90%. Due to the vertical emittance of the primary beam and an eventual shift of the cathode with respect to the mean vertical position of the beam, there was a small probability that both fission fragments passed through the same half of the double IC. Additionally, fission fragments moving very close to the cathode had less active volume to ionise and produced less electron-ion pairs, causing some additional losses.

## 4. Experimental results

The experimental set-up described in section 3 allowed measuring the total fission cross sections of $^{238}$U projectiles at 1 $A$ GeV on C, $(CH_2)_n$, Cu, and Pb targets. Moreover, for the reaction of $^{238}$U at 1 $A$ GeV on $(CH_2)_n$ we determined the partial fission cross sections, that is, the cross section as a function of the fissioning element, and the widths of the partial fission-fragment charge distributions, namely, the widths of the charge distributions of the fission fragments that result from a given fissioning element. The analysis procedure used to determine all these observables is thoroughly described in reference [42]. In sections 4.1 and 4.2 we recall the essentials, only.

In figure 4, a cluster plot with the energy-loss signal in the upper part of the double IC against the energy-loss signal in the lower part is shown. The fission events are included in the displayed window and populate the central peak, while the residues produced in fragmentation reactions and central collisions occupy the edges of the spectrum. As mentioned before, due to the limited efficiency of the double IC the window of figure 4 contains around 90% of the fission events. The total number of fission events could be reconstructed by combining the information of the two MUSICs and the double IC.

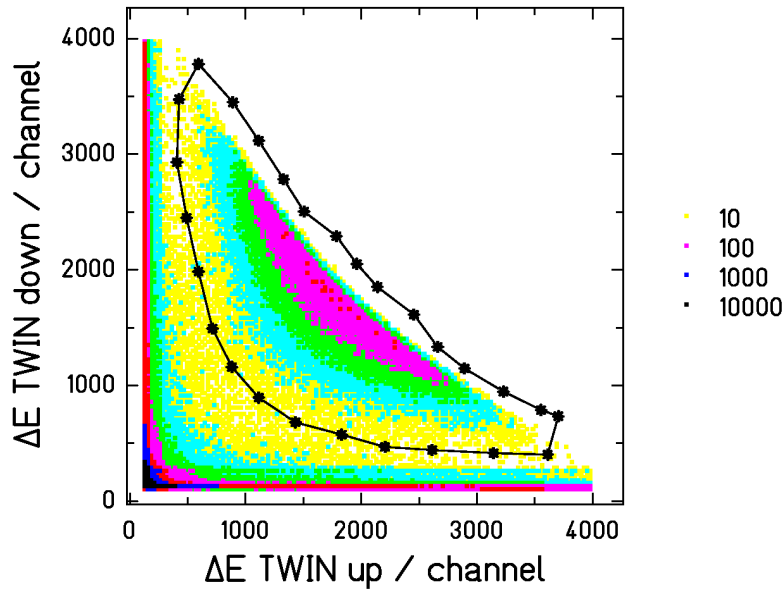

**Figure 4:** Energy-loss signal in the lower part of the double IC versus the energy-loss signal in the upper part of the double IC for the reaction of $^{238}$U (1 $A$ GeV) on $(CH_2)_n$. The window defines the fission events.

### 4.1. Partial fission cross sections

At relativistic energies, the velocity of the fission fragments is close to the velocity of the projectile. In the velocity range of the present experiment, the energy deposition is close to its minimum as a function of velocity and thus it is almost independent of velocity. As the fission residues are fully stripped, the energy loss is directly proportional to the square of the atomic number $Z$. Thus, the double IC allows for determining the nuclear charges of both fission residues, $Z_1$ and $Z_2$. Since they are neutron rich and since their excitation energy is low, the probability for emitting protons after scission is small. Therefore, the sum $Z_1+Z_2$ of the charges of the two fission fragments is an interesting quantity, because it is not very different from the charge of the fissioning nucleus.

Figure 4 shows that the limit between fission and other processes in the region of the lightest fission fragments is somewhat uncertain, while for the heaviest fission fragments the fission peak is well defined. The contribution of fission fragments arising from fissioning elements with charges larger than the projectile charge 92 is approximately 2%. This sharp upper edge of the fission peak was used as reference point for assigning nuclear charges to the energy-loss signals.

Once the charge calibration is done, the fission peak shown in figure 4 transforms into the spectrum of figure 5. We can clearly distinguish the diagonal lines that correspond to fission events for which the sum of the charges $Z_1+Z_2$ is constant. Each of these lines represents the possible combinations of charge splits associated to one fissioning element. From these considerations we conclude that fissioning elements ranging from charge 93 to approximately charge 70 contribute to the total fission cross section of $^{238}$U at 1 $A$ GeV on $(CH_2)_n$. The yield associated to every fissioning element can be better observed representing the number of fission events as a function of $Z_1+Z_2$ as shown in figure 6a). A least-squares Gaussian fit to the individual peaks of this spectrum was performed. The areas of the Gaussians as a function of their centers yield the curve of figure 6b).

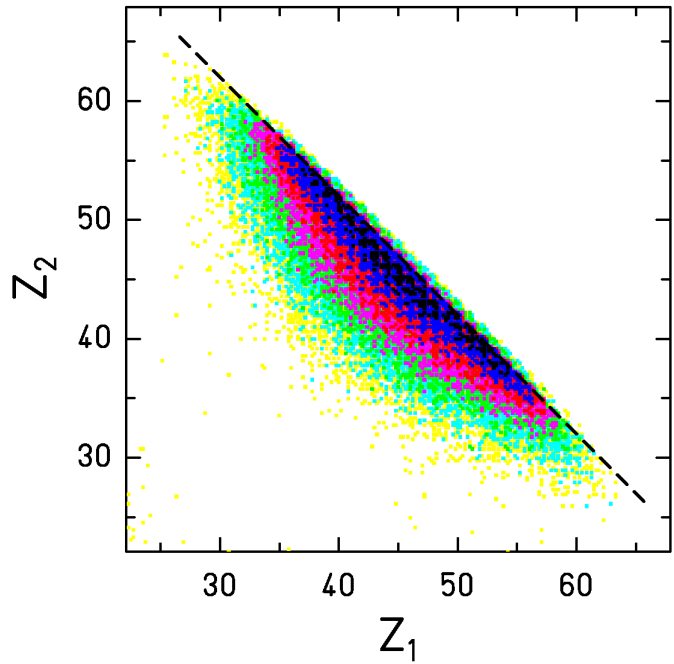

**Figure 5:** Nuclear charge of the fission fragment passing through the lower part of the double IC as a function of the nuclear charge of the complementary fragment passing through the upper part of the double IC for the reaction $^{238}$U (1 $A$ GeV) + $(CH_2)_n$. The dashed line marks the fission events with $Z_1+Z_2 = 92$.

On the path from the target to the double ionisation chamber, the fission fragments have to pass several layers of matter, the most important one being about 2.5 meters of air. On this path, the fission residues may lose some protons in secondary fragmentation reactions. As a consequence, the corresponding fission event will be attributed to a lighter fissioning nucleus $Z_1+Z_2$ than it actually corresponds to. Thus, the yields related to the heaviest fissioning nuclei are underestimated whereas those related to the lightest fissioning nuclei are overestimated. Hence, before transforming the yields of figure 6b) into cross sections, we have to consider that the shape of this spectrum is affected by the secondary reactions of the fission residues on their way to the double IC. We have corrected for this effect by following an iterative process in which the charge-changing cross sections of the fission fragments in the different layers of the set-up were calculated with EPAX [43]. For the $H_2$ of the $(CH_2)_n$-target a fast simplified version of the INCL3 code [44, 45] was employed (see appendix of reference [46]). Figure 7 shows a comparison between the experimental yields before (dashed line) and after (full line) the correction for secondary reactions. From this comparison it follows that the effect of secondary reactions is very small.

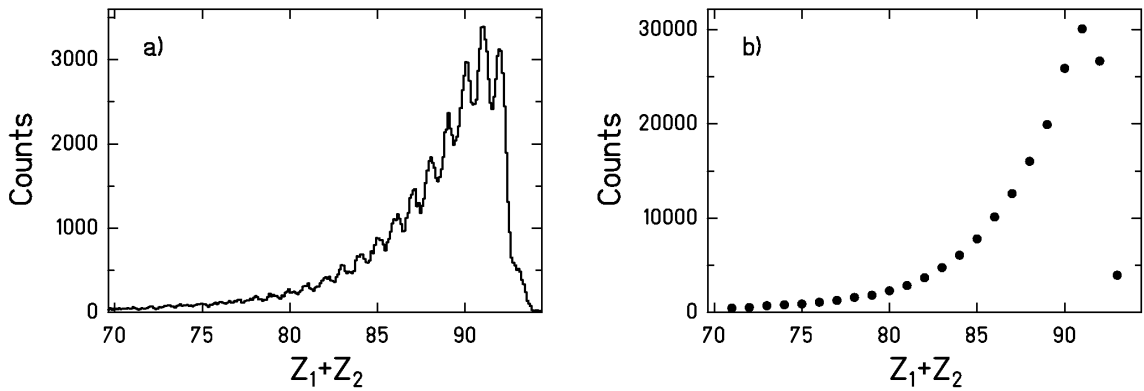

**Figure 6:** Number of fission events as a function of the sum of the nuclear charges of the two fission fragments, $Z_1+Z_2$, for the reaction of $^{238}$U (1 $A$ GeV) + $(CH_2)_n$. a) Raw data after the charge calibration. b) Result of representing the areas of the Gaussian fits to the peaks of part a) as a function of the centers of the Gaussians. The statistical errors of the data points are smaller than the symbols used.

Up to now, we considered only those fission events which are included in the fission window defined in figure 4. The fission events not included in this window are distributed in the same proportion along the whole $Z_1+Z_2$ range. Therefore, we could correct for these losses by normalizing the spectrum of figure 7 (full line) to the total fission cross section determined in [42]. In addition to the uncertainties of the individual partial yields due to the counting statistics and the uncertainties of the corrections for secondary reactions, the uncertainty of the total fission cross section has been added to obtain the final uncertainties of the partial fission cross sections.

Figure 7 reflects how the yields decrease with decreasing charge of the fissioning nucleus. Several effects define this trend. On the one hand, the fission barriers increase with decreasing charge of the fissioning nucleus. On the other hand, the light fissioning nuclei result from more central collisions, which are less probable. Nevertheless, as we will see below, the partial fission cross sections are expected to depend on dissipation as well. A qualitative and quantitative analysis of this dependence is presented in section 5.

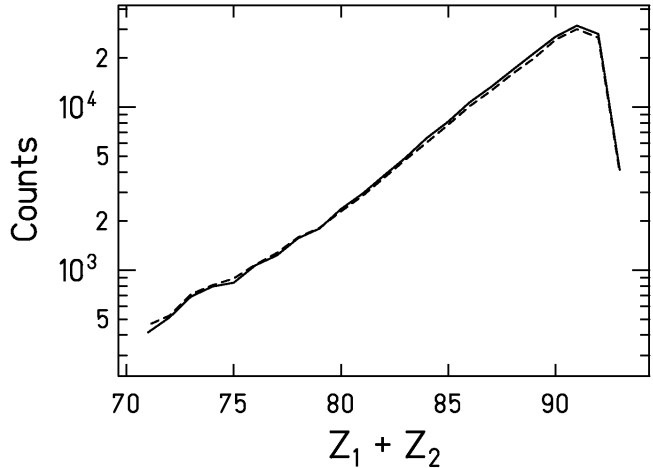

**Figure 7:** Experimental yields (dashed line) for the reaction of $^{238}$U (1 $A$ GeV) + $(CH_2)_n$ in comparison with the same data corrected for secondary reactions (full line). For the seek of clarity, the error bars have not been included.

**4.2. Widths of the partial charge distributions of the fission fragments**

The charge distributions of the fission fragments for a fixed fissioning element $Z_1+Z_2$ produced in the reaction of $^{238}$U at 1 $A$ GeV on $(CH_2)_n$ can be obtained by projecting the diagonal lines of figure 5 either on the horizontal or the vertical axis. Figure 8 shows three examples: the case of $Z_1+Z_2 = 83$, in part a), the case of $Z_1+Z_2 = 90$, in part b) and the case $Z_1+Z_2 = 92$, in part c). The asymmetric charge distribution of the fissioning nucleus with charge $Z_1+Z_2 = 92$ is characteristic for the contribution of low-energy fission. Here, the excitation energy is induced by very peripheral nuclear collisions where none or only few neutrons are abraded and, to a small part, by the electromagnetic interaction with the target nuclei. For lighter fissioning nuclei, the impact parameter becomes smaller and the induced excitation energy increases. Hence, the influence of the shell effects is increasingly attenuated, and the charge distributions become symmetric. In fact, as shown in figure 8b), shell effects almost vanished already for $Z_1+Z_2 = 90$. As will be explained in the next section, when shell effects are washed out, the variance of the charge distribution scales with the temperature at the saddle point. Therefore in our analysis of the charge widths we will only consider symmetric distributions. To determine these widths, Gaussian fits to the charge distributions based on the method of the maximum likelihood were performed, assuming the data to be Poisson distributed. The result of the fit for the cases of $Z_1+Z_2=83$ and $Z_1+Z_2=90$ is shown by the full lines of figures 8a) and 8b), respectively. The lighter the fissioning nucleus, the larger is the contamination of the tails of the charge distribution by residues originating from other reaction mechanisms. To avoid that the shape of the distribution is falsified by this effect, the tails of the distribution do not enter into the Gaussian fits, which only consider the central part of the spectrum, clearly attributed to fission. For instance, in figure 8a), the central part between $Z = 27$ and $Z = 56$ closely follows the Gaussian distribution of the fission process. However, the tails tend to be higher due to contributions from other reaction mechanisms, until they are cut at $Z = 21$ and $Z = 62$ by the two-dimensional window shown in figure 4.

The standard deviations of the charge distributions are represented in figure 9 as a function of $Z_1+Z_2$. The error bars represent the uncertainties of the fits. A model calculation revealed that the standard deviations do not change when the correction for secondary reactions of the fission fragments is considered.

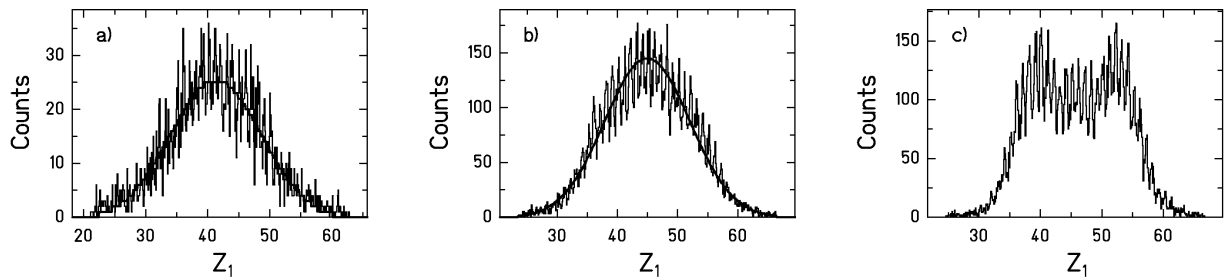

**Figure 8:** Fission-fragment element distribution for the fissioning nucleus $Z_1+Z_2 = 83$, figure a) and for the fissioning nucleus $Z_1+Z_2 = 90$, figure b). The full lines are the result of Gaussian fits. c) Fission-fragment element distribution for the fissioning nucleus $Z_1+Z_2 = 92$.

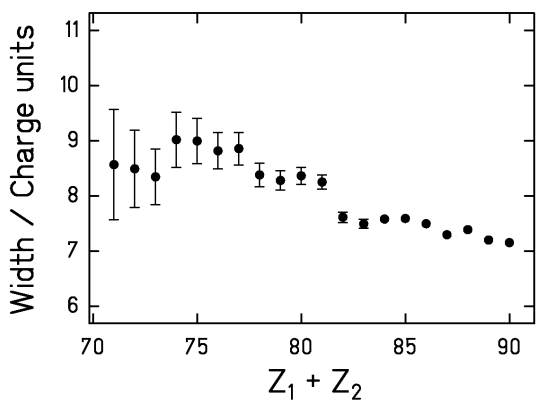

**Figure 9:** Standard deviations of the charge distributions as a function of the charge sum of the fission fragments $Z_1+Z_2$ for the reaction of $^{238}$U at 1 $A$ GeV on $(CH_2)_n$.

## 5. Interpretation of the experimental results

Fission and evaporation-residue cross sections are strongly sensitive to dissipation in the small deformation regime. They have been studied up to now with different experimental approaches. In most experiments, these observables result from the contribution of many fissioning nuclei of different nature: mass, charge, excitation energy and angular momentum. The involvement of such large variety of systems might cause a net compensation that hides the real tendencies. Therefore, these observables do not permit for constraining the parameters of the theoretical codes in an optimum way. Providing additional data that reveal the distribution of fissioning nuclei or that are directly sensitive to the excitation energy at fission allows for a more critical test of the theoretical models. In this spirit, a recent work [19, 20] introduced the angular-momentum distribution of the evaporation residues as an additional observable sensitive to dissipation inside the saddle, since the mean value of the angular-momentum distribution shifts to larger values with increasing dissipation. In the next section we will show in a qualitative way why the total fission cross sections as a function of the target charge, the partial fission cross sections and the partial widths of the charge distributions of the fission fragments are sensitive to dissipation. A quantitative conclusion on the strength of dissipation can be obtained from a direct comparison of the experimental data with a nuclear-reaction code. This will be illustrated in section 5.3.

**5.1. New experimental signatures for transient effects**

We would like to start this section by mentioning that transient effects may only be observable in a limited excitation-energy range of the fissioning system. As the decay time for particle emission decreases exponentially with increasing excitation energy, there exists a threshold excitation energy from which the fission process starts to be sensitive to the dynamical delay, even without considering a temperature-dependent dissipation coefficient. This can be seen in the calculation of figure 10 that illustrates the dependence of the mean value of the transient time $<\tau_{tran}>$ (dashed line) on the excitation energy at fission together with the same dependence of the mean fission time $<\tau_{sad}>$, i.e. the average time that the system needs to pass the fission barrier (full line). The transient time has been calculated according to equation (4), and the mean fission time is the calculated time span from the abrasion up to fission. The time of the break-up stage is considered to be negligible.

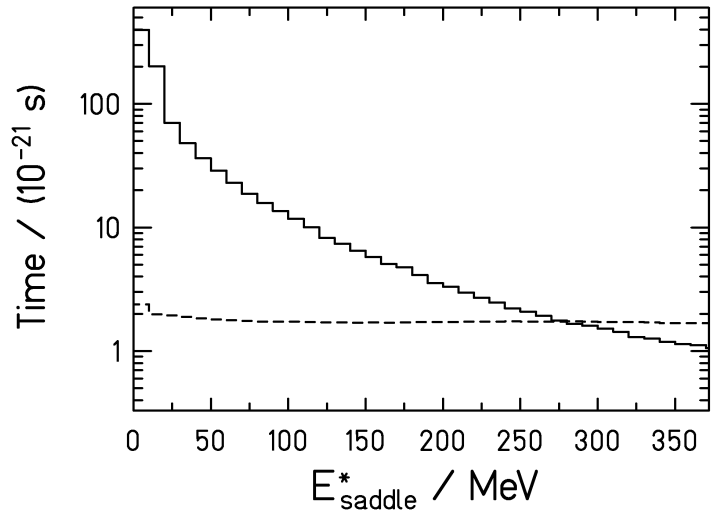

**Figure 10:** Calculation representing the mean time $<\tau_{sad}>$ the system needs to cross the saddle point (full line) as a function of the excitation energy at saddle in comparison with the average transient time $<\tau_{tran}>$ (dashed line) as a function of the excitation energy at saddle. The calculation was performed for the reaction $^{238}$U (1 $A$ GeV) + Pb with $\beta = 2 \cdot 10^{21}$ s$^{-1}$. It includes the break-up stage, and $\Gamma_f(t)$ follows the analytical solution of the FPE [28, 29].

Figure 10 shows that the mean fission time $<\tau_{sad}>$ decreases strongly with increasing excitation energy, while the transient time shows only a very weak dependence. For low excitation energies, the mean fission time is several orders of magnitude larger than the transient time. At excitation energies around 200 MeV, however, $<\tau_{tran}>$ approaches $<\tau_{sad}>$ within a factor of two. For the highest energies, the mean fission time is even smaller than the transient time due to the strongly decreasing stationary fission-decay time $\hbar/\Gamma_K$. Figure 10 reveals that for nuclei with excitation energies lower than about 100 MeV the transient time is still a too tiny effect to have observable influence on the fission decay width. This should be kept in mind when looking for experimental signatures that are sensitive to transient effect.

In our experiment, the nuclear charges of both fission products from the interaction of the $^{238}$U relativistic beam with different targets were determined. This information permits to extract the fission cross sections as a function of the target size. If the excited nucleus experiences a dynamical hindrance on its way to the saddle point, contrary to what is assumed in the Bohr-Wheeler transition-state model, the fission channel will be initially closed in the deexcitation process. Hence, if the particle-decay time is shorter than the dynamical fission delay, the nucleus will first release excitation energy by evaporation. Due to the loss of excitation energy, the fission probability of the resulting system after the dynamical delay will be considerably smaller than the one of the original compound nucleus. Consequently, the fission cross sections will be reduced compared to the transition-state-model predictions. In addition, since the target has an influence on the mass and the excitation energy of the residues produced directly after the collision, dissipation might affect the number of systems that fission for each target in a different way.

As stated before, the sum of the nuclear charges of the fission residues, $Z_1 + Z_2$, corresponds closely to the charge of the fissioning nucleus. Although there might be a small number of charged particles emitted after the fragmentation, the charge of the fissioning element grows with the charge of the prefragment and, hence, it gives an indication of the centrality of the collision. Low values of $Z_1+Z_2$ imply small impact parameters and large excitation energies induced by the nuclear collision.

If the excitation energy of the nucleus is higher than the threshold excitation energy at which the statistical decay time and the transient time $\tau_{tran}$ become comparable, the fission competition is delayed with respect to evaporation due to transient effects. Therefore, for the lightest fissioning nuclei, i. e. the lowest values of $Z_1+Z_2$, the existence of a transient time would imply a considerable reduction of the fission cross section compared to the predictions of the transition-state model.

The width of the charge distribution[e] of the fission fragments was used already in reference [18] to determine the temperature of the fissioning system at saddle. The relation between the width of the charge distribution and the saddle-point temperature $T_{sad}$ relies on an empirical systematic which relates the variance of the mass distribution $\sigma_A^2$ to $T_{sad}$ that reads

---

[e] The charge distribution considered here is the integral distribution of elements produced. It should not be confounded with the charge distribution of isobaric chains (A=const.), also often dealt with in fission.

$$\sigma_A^2 = \frac{T_{sad}}{C_A} \tag{5}$$

where $C_A$ is a constant that depends on the fissility $Z^2/A$ of the fissioning nucleus. Due to the almost constant mass-to-charge ratio of the fission fragments, the charge and the mass of the fission fragments are strongly correlated, and thus a linear relation between $\sigma_A^2$ and $T_{sad}$ directly implies a linear relation between $\sigma_Z^2$ and $T_{sad}$ as well. Thus, for the lower values of $Z_1+Z_2$, where the initial excitation energy is larger than the threshold excitation energy, the nucleus will evaporate particles on its path to fission, and $T_{sad}$ will be smaller than the initial temperature. Consequently, the corresponding charge distributions will be narrower than the ones predicted by the transition-state model. The quantity $C_Z$ that connects $\sigma_Z^2$ with $T_{sad}$ is related to $C_A$ as follows

$$C_Z = \frac{A_{fiss}^2}{Z_{fiss}^2} C_A \tag{6}$$

where $A_{fiss}$ and $Z_{fiss}$ represent the mass and the nuclear charge of the fissioning nucleus, respectively.

Rusanov et al. [47] analysed the experimental mass-energy distributions of fragments produced in the fission of various nuclei with fissilities $Z^2/A \geq 32$ and excitation energies between 40 and 150 MeV. Their study revealed a considerable enhancement of the variances $\sigma_A^2$ with the angular momentum. Furthermore, the relation between $C_A$ and the fissility parameter was intensively studied. In a later work [48], the experimental mass distributions of the fission products of proton- and alpha-induced reactions were investigated. By means of a fitting procedure, the values of the parameter $C_A$ for nuclei within a range of fissilities $Z^2/A$ from about 28 to 44 and excitation energies from approximately 5 to 25 MeV were determined. Including the data for larger fissilities of reference [47], corrected for angular-momentum effects, a parameterisation of the relation between $C_A$ and the fissility was deduced.

According to the transition-state model, the constant $C_A$ is proportional to the stiffness of the liquid-drop potential at the saddle point with respect to the mass-asymmetric deformation. In the frame of the transition-state model, the mass yields $Y(A)$ follow the expression

$$Y(A) \propto \exp\left[2\sqrt{a_f(E^* - B_f(A))}\right] \tag{7}$$

where $A$ is the mass of the fragments, $a_f$ is the level-density parameter at saddle, and $E^*$ is the excitation energy of the fissioning nucleus. The conditional fission barrier $B_f(A)$ corresponding to the fragment mass $A$ can be described by the liquid-drop model including shell effects by

$$B_f(A) = (B_f^{LD} - W_g) + K_A\left(A - \frac{A_{fiss}}{2}\right)^2 + W_f(A) \tag{8}$$

where $B_f^{LD}$ is the liquid-drop fission barrier, $W_g$ and $W_f(A)$ are the shell corrections in the ground state and in the transition state, respectively. The stiffness parameter $K_A$ of the liquid drop depends on the mass asymmetry at saddle as follows:

$$K_A = \frac{1}{2} \frac{d^2 V(A)}{dA^2}\bigg|_{A=A_{fiss}/2} \qquad (9)$$

Inserting equation (8) into equation (7) and taking into account that at high excitation energies shell effects can be neglected, the mass yield $Y(A)$ is approximated by a Gaussian with the standard deviation

$$2\sigma_A^2 = \frac{\sqrt{E_{sad}^*/a_f}}{K_A} = \frac{T_{sad}}{K_A} \qquad (10)$$

where $E_{sad}^*$ is the excitation energy of the fissioning nucleus above the fission barrier. Comparing equation (10) with equation (5), it follows that $C_A = 2K_A$.

Although the arguments that lead to equation (10) were based on the statistical model, the linearity between the variance of the mass distribution and the temperature at saddle has been confirmed by dynamical calculations [49]. However, in this more complete picture the quantity $C_A$ cannot be just interpreted as the stiffness of the liquid drop potential at saddle because it includes as well the dynamical effects that influence the charge distribution on the way from saddle to scission. In reference [49], two-dimensional Langevin calculations were used to investigate the mass distribution of the fragments produced in the fission process of compound nuclei within the fissility range $20 < Z^2/A < 40$. For the compound nucleus $^{205}$At, the linear dependence of $\sigma_A^2$ on $T_{sad}$ was confirmed for saddle-point temperatures reaching up to 2.4 MeV by this calculation. This gives an indication that the parameterisation of reference [48] is valid also at higher excitation energies. In the quantitative analysis of the widths of the experimental charge distributions, we will take the appropriate value of $C_Z$ for each fissioning nucleus from the parameterisation introduced in reference [48].

### 5.2. The nuclear-reaction code

The nuclear-reaction code used is an extended version of the abrasion-ablation Monte-Carlo code ABRABLA [31, 45]. It consists of three stages: In the first stage, the interaction of the two reaction partners is described by the abrasion model. If highly excited, the resulting system then eventually undergoes a break-up process. Finally, the ablation stage is considered where the sequential decay takes place.

*5.2.1. Abrasion stage*

The abrasion model [30] is well suited for describing the properties of the projectile fragment after peripheral collisions. The basic idea of this model is that at relativistic energies ($> 100\ A$ MeV) the relative velocity of the reaction partners is large compared to the Fermi velocity of the nucleons in the potential well. In addition, the associated de-Broglie wavelength of the projectile is of the order of the size of the nucleons. Thus, in the overlap zone between the projectile and the target many nucleon-nucleon collisions take place, while the nucleons in the non-overlapping region are only little disturbed.

Depending on the impact parameter, a distribution of projectile fragments with different masses and nuclear charges are formed by the abrasion process. The mass loss of the projectile, respectively the target, can be determined geometrically by integrating the overlapping volume of the two reaction partners. For a given mass loss, the *N/Z* distribution is calculated assuming that every nucleon removed has a statistical chance to be a neutron or a proton as determined by the neutron-to-proton ratio of the precursor nucleus. That results in a hypergeometrical distribution [50] centred at the *N/Z* ratio of the projectile. The angular-momentum distribution [33] of the projectile fragments is given in analogy to Goldhaber's description for the linear-momentum distribution of the projectile residues, which has proven to be in good agreement with experiment [51]. According to this idea, the angular-momentum distribution is defined by the initial angular momenta of the nucleons removed. In the same way, the excitation-energy distribution of the projectile residues is determined by the sum of the energies of the holes in the single-particle scheme of the initial nucleus which are formed due to the removal of nucleons. Including final-state interactions derived from measured isotopic production cross sections [32], an average excitation energy of 27 MeV per nucleon abraded is induced. This value is in agreement with predictions for peripheral collisions based on BUU calculations [52].

*5.2.2. Break-up stage*

Mid-peripheral heavy-ion collisions allow producing nuclei with excitation energies that are far beyond the onset of multifragmentation [53]. Nuclear dynamics in this range of excitation energies is the subject of current research. As presented in reference [54, 36], the analysis of the isotopic distributions of heavy projectile fragments from the reactions of a $^{238}$U beam in a lead target and a titanium target gave evidence that the initial temperature of the last stage of the reaction, the evaporation cascade, is limited to a universal upper value of approximately 5.5 MeV. The interpretation of this effect relies on the onset of a simultaneous-break-up process for systems whose temperature after abrasion is larger than 5.5 MeV. In ABRABLA, the simultaneous break-up stage has been modelled in a quite rough way that, however, shows a good agreement with the experimental data. If the temperature after abrasion exceeds the freeze-out temperature of 5.5 MeV, the additional energy is used for the formation of clusters and the simultaneous emission of these clusters and several nucleons. The number of protons and neutrons emitted in the break-up stage is assumed to conserve the *N*-over-*Z* ratio of the projectile spectator, and an amount of about 20 MeV per nuclear mass unit emitted is released. This last quantity is still under investigation. Nevertheless, its effect on the results is very small. The break-up stage is assumed to be very fast, and thus the fission collective degree of freedom is not excited. The resulting piece left of the projectile spectator then undergoes the sequential decay.

*5.2.3. Ablation stage*

We assume that at initial temperatures exceeding 5.5 MeV, after the simultaneous break-up and for lower temperatures directly after abrasion, the intrinsic degrees of freedom of the projectile residue reach thermal equilibrium very fast. The following ablation stage describes the sequential deexcitation of the projectile fragment by particle evaporation and/or fission. The decay widths for particle emission are obtained from the statistical model; for instance, the neutron width is given by [55]:

$$\Gamma_n(E) = \frac{1}{2\pi\rho(E)} \frac{4m_n R^2}{\hbar^2} T_r^2 \rho_r(E - S_n) \quad (11)$$

$\rho(E)$ is the level density of the compound nucleus, $S_n$ is the neutron separation energy, $m_n$ is the neutron mass, $R$ is the nuclear radius, $\rho_r(E)$ is the level density of the daughter nucleus after neutron emission and $T_r$ is the temperature of the residual daughter nucleus after neutron emission. The proton- and alpha decay widths, $\Gamma_p$ and $\Gamma_\alpha$, follow an analogue formula, but the level densities of the daughter nuclei, $\rho_r(E-S_p-B_p^{eff})$ and $\rho_\alpha(E-S_\alpha-B_\alpha^{eff})$, are shifted by the respective effective Coulomb-barriers, $B_p^{eff}$ and $B_\alpha^{eff}$.

As discussed in the introduction, dissipation causes an initial suppression of the fission width. Later, this width grows and finally reaches its asymptotic value given by the product of the Kramers factor (equation 1) and the Bohr-Wheeler fission decay width $\Gamma_f^{BW}$, which is determined according to the transition-state model [2], see also [55]:

$$\Gamma_f^{BW} = \frac{1}{2\pi\rho(E)} T_{sad} \rho_{sad}(E - B_f) \quad (12)$$

The level density of the fissioning nucleus at saddle is $\rho_{sad}(E)$, and the level density of the compound nucleus is $\rho(E)$. The temperature of the nucleus at saddle is $T_{sad}$. The implementation of the time dependence of the fission width is crucial in order to obtain reliable results. This is a rather complicated task, and in literature one mostly finds two rather crude approximations, a step function [56] that sets in at time $\tau_{tran}$:

$$\Gamma_f(t) = \begin{cases} 0, t < \tau_{tran} \\ \Gamma_f^K, t \geq \tau_{tran} \end{cases} \quad (13)$$

and an exponential in-growth function [57]:

$$\Gamma_f(t) = \Gamma_f^K \cdot \{1 - \exp(-t/\tau)\} \quad (14)$$

where $\tau = \tau_{tran}/2.3$ and $\Gamma_f^K$ is the Bohr-Wheeler fission width given by equation (12), multiplied by the Kramers factor $K$ of equation (1). In references [28, 29] we have intensively studied the effect of the description of the time dependence of $\Gamma_f$ on the deduced magnitude of $\beta$, and we have developed a realistic description for $\Gamma_f(t)$ based on the analytical solution of the FPE. In order to investigate how the time-dependent shape of the fission-decay width influences the analysis, these three descriptions of $\Gamma_f(t)$ have been implemented in the ABRABLA code.

Besides the treatment of the dissipation effects, the most critical ingredients that define the fission width are the ratio of the level-density parameters $a_f/a_n$, and the fission barriers $B_f$. The ratio $a_f/a_n$ is calculated considering volume and surface dependencies as proposed in reference [58] according to the expression:

$$a = \alpha_v A + \alpha_s A^{2/3} B_s \quad (15)$$

where $\alpha_v$ and $\alpha_s$ are the coefficients of the volume and surface components of the single-particle level densities, respectively, with the values $\alpha_v = 0.073$ MeV$^{-1}$ and $\alpha_s = 0.095$ MeV$^{-1}$. It has been shown recently, that this parameterisation is in good agreement with expectations from the finite-range liquid-drop model [59]. $B_s$ is the ratio of the surface of the deformed nucleus relative to the surface of a spherical nucleus. Its value is taken from reference [60]. The angular-momentum-dependent fission barriers are taken from the finite-range liquid-drop model predictions of Sierk [61]. As demonstrated in reference [18], a recent experimental determination of these parameters by K. X. Jing and co-workers [16], based on the measurement of cumulative fission probabilities of neighbouring isotopes, is in very good agreement with these theoretical predictions.

### 5.3. Comparison of the experimental data with model calculations

Let us examine now the sensitivity of the experimental observables to the value of the reduced dissipation coefficient $\beta$. Since we have shown [28] that the approximation for the fission decay we have developed is close to the numerical solution of the FPE, the subsequent calculations are performed with this analytical expression. In Figure 11 we consider the partial fission cross sections and the widths of the charge distributions for the reaction $^{238}$U (1 $A$ GeV) + (CH$_2$)$_n$. The experimental data (full dots) are compared with the predictions of the transition-state model (dashed line) and with several calculations obtained for three values of $\beta$, corresponding to three different dissipation regimes. In all calculations, the interactions with carbon were calculated with the ABRABLA code, while the contribution given by the hydrogen part of the (CH$_2$)$_n$ target has been determined with a fast simplified version of INCL3 [44, 45, 46] combined with the same evaporation code as used in ABRABLA. The staggering of the curves and the strong decrease of the dashed-dotted curve below $Z_1 + Z_2 \approx 78$ in Figure 11b) are due to statistical fluctuations of the Monte-Carlo calculations.

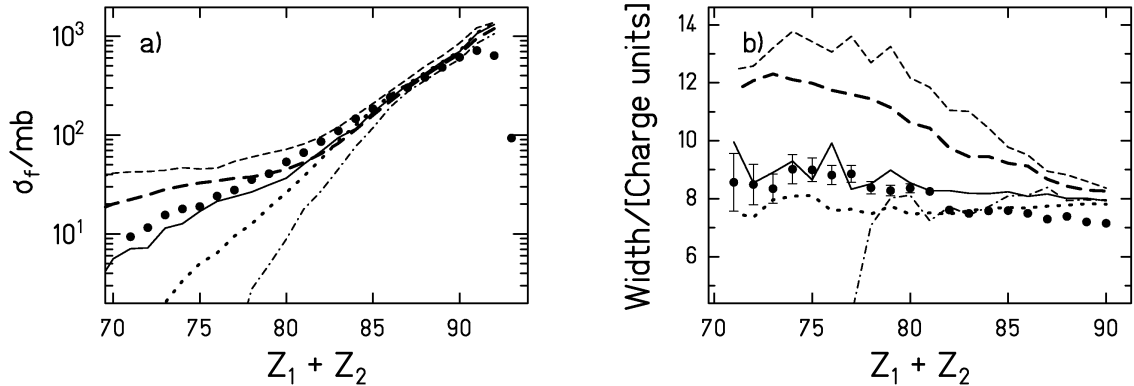

**Figure 11:** Experimental partial fission cross sections (full dots part a) and widths of the charge distributions of the fission fragments (full dots part b) obtained for the reaction $^{238}$U (1 $A$ GeV) + (CH$_2$)$_n$ in comparison with various ABRABLA calculations. Error bars smaller than the symbols are not shown. The thin dashed lines and the thick dashed lines correspond to the transition-state model and Kramers stationary solution with $\beta = 6 \cdot 10^{21}$ s$^{-1}$, respectively. The other calculations are performed using the fission width that follows from the analytical solution of the FPE [28, 29] with $\beta = 2 \cdot 10^{21}$ s$^{-1}$ (full lines), $\beta = 0.5 \cdot 10^{21}$ s$^{-1}$ (dotted lines) and $\beta = 5 \cdot 10^{21}$ s$^{-1}$ (dashed-dotted lines).

As expected, both observables are over-estimated by the transition-state model confirming their sensitivity to dissipation. But also the Kramers stationary solution deviates from the data. In particular the width of the charge distribution is strongly overestimated. For the partial fission cross sections (figure 11a) as well as for the partial widths of the charge distributions (figure 11b), the best description is given by the full line that corresponds to $\beta = 2\cdot10^{21}$s$^{-1}$. Such value of $\beta$ corresponds to the shortest transient time $\tau_{tran}$. This means that the systems studied are critically damped: The coupling between the collective and the intrinsic degrees of freedom leads to the fastest possible spread of the probability distribution in deformation space. As shown in figure 11, any other value of $\beta$ above or below the critical damping will result in lower cross sections and narrower distributions. This is due to the increase of the transient time $\tau_{tran}$ for all values of $\beta$ which are smaller or larger than $2\omega_g$, see equation (4).

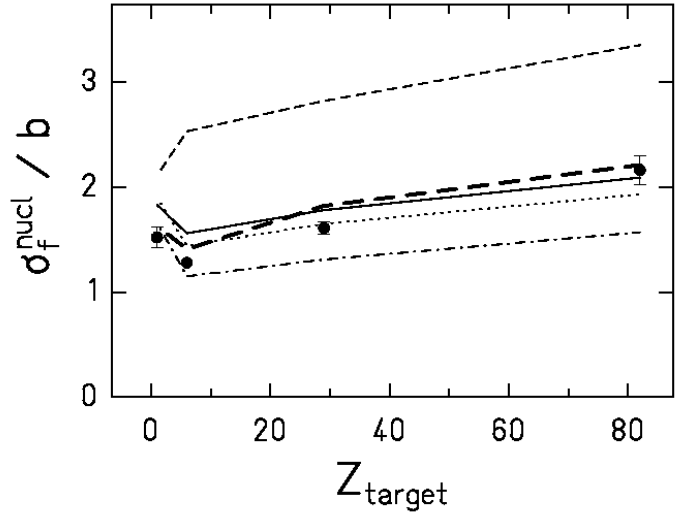

**Figure 12:** Total nuclear fission cross sections of $^{238}$U (1 $A$ GeV) as a function of the target nuclear charge. The experimental data are represented by the full dots. The short-dashed line is a calculation according to the transition-state model, while the long-dashed line represents the calculation with the Kramers fission width using $\beta = 6\cdot10^{21}$ s$^{-1}$. The other calculations are performed using the fission width that follows from the analytical solution of the FPE [28, 29]. The full line corresponds to $\beta = 2\cdot10^{21}$s$^{-1}$, the dotted line corresponds to $\beta = 0.5\cdot10^{21}$s$^{-1}$ and the dashed-dotted line to $\beta = 5\cdot10^{21}$s$^{-1}$.

The dependence of the total nuclear fission cross sections on the charge of the target is portrayed by the full dots in figure 12 together with several ABRABLA calculations. The nuclear fission cross sections are obtained subtracting the electromagnetic component as calculated in reference [62] from the total fission cross sections determined in [42]. Since the total reaction probability increases with the mass of the target, one would expect that the fission cross sections grow with the charge of the target as well. However, figure 12 shows that the nuclear fission cross section is smaller for carbon than for hydrogen. As before, the calculation based on the transition-state model (dashed line) overestimates the experimental values and does not reproduce the minimum of the cross section for $Z_{target} = 6$. Although none of the calculations that include dissipation provides a good quantitative description of the whole set of data, the existence of the minimum for $Z_{target} = 6$ is only reproduced when dissipation is considered. (In fact, the calculation with the Kramers fission width and $\beta = 6\cdot10^{21}$ s$^{-1}$ reproduces the data best, but as we saw this calculation fails to describe the two other experimental signatures, see Figure 11.) We think that the remaining failure to reproduce the target dependence of the cross section is due to an unexplained shortcoming of the abrasion model that was noticed already by Rubehn et al. [62]. Nevertheless, the fact that the minimum of the cross section is found for carbon and not for hydrogen has its explanation in dissipation. The fragmentation of $^{238}$U at 1 $A$ GeV in hydrogen leads to the production of heavy projectile-like residues with low excitation energies and low fission barriers. These nuclei are not sensitive to the dynamical delay induced by dissipation, and most of them fission. On the other hand, when the $^{238}$U projectiles at 1 $A$ GeV react with a carbon target, the number of heavy projectile-like products decreases and the residue distribution extends to lighter nuclei with higher excitation energies. If there would be no dynamical delay, great part of these nuclei would fission, and the total fission cross section would increase with respect to the hydrogen case. However, if dissipation is considered, the statistical decay time of these lighter highly excited nuclei is similar or even shorter than the dynamical delay, and fission is suppressed leading to lower cross sections.

As discussed in reference [29], there exist solid theoretical arguments why the exponential-like in-growth function for describing $\Gamma_f(t)$ does not express correctly the increase of the fission decay width as a function of time. This function does not reproduce the initial suppression of fission as shown up in the solution of the FPE and therefore allows for fission at very high excitation energies. Thus, it is to expect that this function is not able to describe properly the experimental data. However, references [28, 63] show that the total nuclear fission cross sections of $^{238}$U (1 $A$ GeV) interacting with different targets can be well reproduced by all the descriptions for $\Gamma_f(t)$, provided that the appropriate value of $\beta$ is used. The exponential function leads to $\beta = 4 \cdot 10^{21}$ s$^{-1}$ and the two other approximations to $\beta = 2 \cdot 10^{21}$ s$^{-1}$. Consequently, we need additional observables that allow tagging the fission events according to the excitation energy for providing experimental evidence on the validity of the different descriptions of the time-dependent fission-decay width. We expect that, for the largest excitation energies, the exponential-like in-growth function leads to results that differ from the measured data. This selection according to the excitation energy can be achieved by considering the charge sum of the fission fragments $Z_1+Z_2$. We already discussed in the former section that the excitation energy of the systems before entering the sequential decay is likely to increase with decreasing $Z_1+Z_2$. However, one has to consider as well that very highly excited prefragments undergo a simultaneous break-up that sets a limit of 5.5 MeV to the temperature of the fissioning nucleus. Thus, one could suspect that for the lightest fissioning nuclei the excitation energy remains constant or even decreases. To clarify the situation, we performed a calculation with ABRABLA that gives the excitation energy of the system right before entering the ablation stage as a function of $Z_1+Z_2$. The result for the carbon[f] target is displayed in figure 13. It can be seen that, initially, the excitation energy increases with decreasing $Z_1+Z_2$, while in the range from $Z_1+Z_2 \approx 78$ to $Z_1+Z_2 \approx 73$ it remains more or less constant with a mean value of approximately 550 MeV, and finally it slowly decreases. Thus, in spite of the simultaneous break-up, the lowest values of $Z_1+Z_2$ are related to excitation energies that are high enough for the statistical decay time to be similar or shorter than the transient time $\tau_{tran}$, and thus the influence of dissipation is expected to be clearly observable.

---

[f] For the case of the (CH$_2$)$_n$ target, we have to add the effect of the hydrogen part that leads mostly to heavy residues and for which the excitation energy induced per removed nucleon is not 27 MeV as in carbon but approximately 50 MeV [J. Cugnon, C. Volant, and S. Vuillier, Nucl. Phys. A 620 (1997) 475.]. Consequently, for the largest values of $Z_1+Z_2$ the spectrum of the correlation between $Z_1+Z_2$ and excitation energy for the (CH$_2$)$_n$ target should be somewhat broader in $E^*$ than the one displayed in figure 13.

Figure 14 shows the same two observables as figure 11. The experimental data (full dots) are compared with three ABRABLA calculations. The dashed line corresponds to the result of using the exponential-like function and $\beta = 4\cdot 10^{21}$ s$^{-1}$, the dotted line was obtained with the step function and $\beta = 2\cdot 10^{21}$ s$^{-1}$, and the full line results from the approximation of the Focker-Planck solution and $\beta = 2\cdot 10^{21}$ s$^{-1}$. These values of $\beta$ are the ones that lead to the best description of the measured total nuclear fission cross sections with each approximation [28, 63]. For both observables, the calculation performed with the fission width derived from the analytical solution of the FPE and the calculation that employs the step function almost coincide over the whole $Z_1+Z_2$ interval. Moreover, in the case of figure 14a), the three calculations agree quite well with each other and with the experimental data for the highest values of $Z_1+Z_2$ but start to differ for the lowest values of $Z_1+Z_2$. In this part of the spectrum, the exponential-like in-growth function overestimates the experimental partial fission cross sections, while the two other functions underestimate them. A slight reduction of the fission barriers [64] at high excitation energies could eventually account for this deviation in the latter two cases. It is not possible to decide from figure 14a), which function $\Gamma_f(t)$ gives the best description of the data. However, in contrast to the partial fission cross sections, the widths of the charge distributions of the fission fragments depicted in figure 14b) indicate a significant disagreement between the calculation done with the exponential-like description and the data. The over prediction of the widths when applying the exponential-like function suggests that this description yields too large excitation energies at saddle. Finally, figure 14a) shows that the model calculations clearly overestimate the partial fission cross sections for $Z_1+Z_2 = 91$ and 92. This is possibly due to a failure of the abrasion model in reproducing very peripheral collisions. However, this discrepancy is not critical for the present discussion, because our conclusions are derived from the fission of light nuclei where they have no influence.

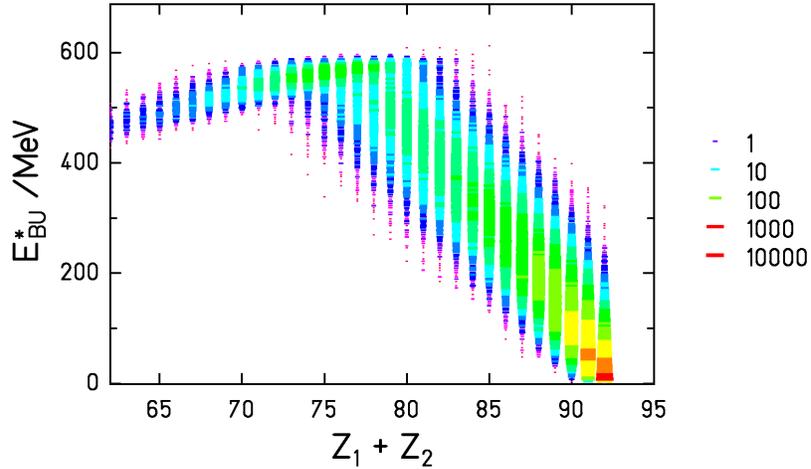

**Figure 13:** Calculation performed with ABRABLA including break-up representing the excitation energy of the prefragment right before entering the ablation stage as a function of the charge sum of the fission fragments. The reaction considered is $^{238}$U (1 $A$ GeV) + C.

All the observables studied here are only sensitive to dissipation in the small-deformation range from the ground state to the saddle point. The value of $\beta = 2\cdot 10^{21}$ s$^{-1}$ resulting from our analysis coincides with the value found in reference [18], where total fission cross sections as well as the widths of the charge distributions and velocities of the fission residues from the reaction Au (800·$A$ MeV) + p are analysed. Other work [16, 19, 65], sensitive to the same deformation range, is consistent as well with our value of $\beta$, although often only upper limits for the transient time or the dissipation strength could be deduced. The value extracted for the reduced dissipation coefficient remains model dependent to a certain degree. Nevertheless, variations of the most critical model parameters by reasonable amount: excitation energy of the prefragments by 30%, freeze-out temperature by 20 % and excitation-energy reduction per mass loss in the break-up stage by a factor of two, led to variations of the deduced transient time well inside the uncertainty range given below. The influence of the ratio of the level-density parameters $a_f/a_n$ has been studied in detail in reference [66] showing that the value of $a_f/a_n$ that gives a good description of our data is the one predicted by Ignatyuk et al. [58]. Although the calculations presented in reference [66] do not include the break-up stage, recent calculations show that the conclusions on $a_f/a_n$ derived there are still valid when the break-up process is considered.

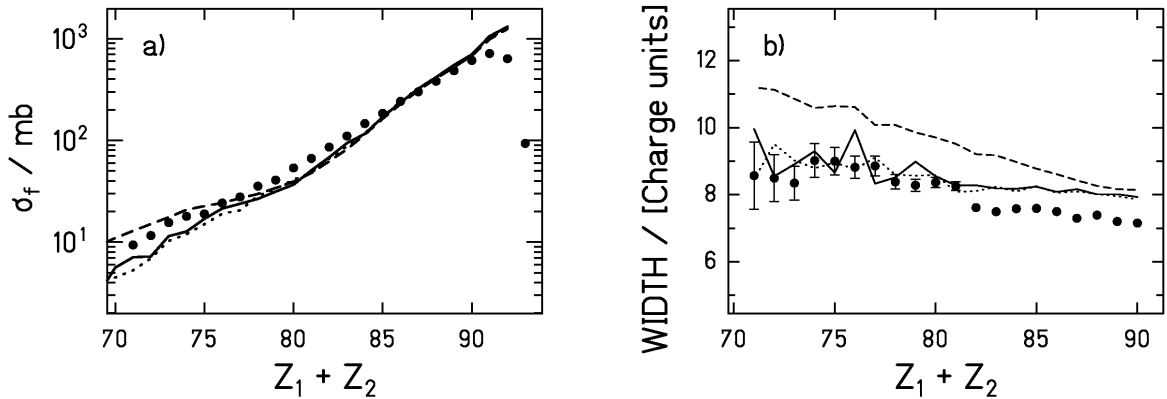

**Figure 14:** Experimental partial fission cross sections (full dots part a) and widths of the charge distributions of the fission fragments (full dots part b) obtained for the reaction $^{238}$U (1 $A$ GeV) + (CH$_2$)$_n$ in comparison with various ABRABLA calculations. The dashed lines correspond to calculations performed with the fission-decay width described by the exponential in-growth function and $\beta = 4 \cdot 10^{21}$s$^{-1}$, the dotted lines are calculations carried out with $\Gamma_f(t)$ approximated by the step function and $\beta = 2 \cdot 10^{21}$s$^{-1}$, and the full lines result from using the fission width that follows the analytical solution of the FPE [28, 29] and $\beta = 2 \cdot 10^{21}$s$^{-1}$.

### 5.4. Transient time

The nuclei that contribute to the fission cross section extend over a broad range of masses, nuclear charges and excitation energies at the saddle point. Hence, according to equation (4) the deduced value of $\beta = 2 \cdot 10^{21}$ s$^{-1}$ corresponds to a distribution of transient times $\tau_{tran}$. In figure 15a) the excitation energy of the fissioning nuclei is represented as a function of the transient time $\tau_{tran}$ determined according to equation (4) assuming $\hbar\omega_{sad} = 1$ MeV [56]. As discussed in [63], the description that reproduces our experimental data implies that transient effects are observed only in those nuclei with excitation energies larger than approximately 150 MeV. Therefore, we only consider the transient time of nuclei with excitation energies at fission above this threshold. The distribution of transient times obtained under the condition $E^*_{sad} > 150$ MeV is shown in figure 15b). A mean value of the transient time of $\tau_{tran} \approx (1.7\pm0.4) \cdot 10^{-21}$ s can be extracted from this curve. Although the distribution illustrated in figure 15b) corresponds to the reaction on the lead target, the mean value of $\tau_{tran}$ is independent of the target considered.

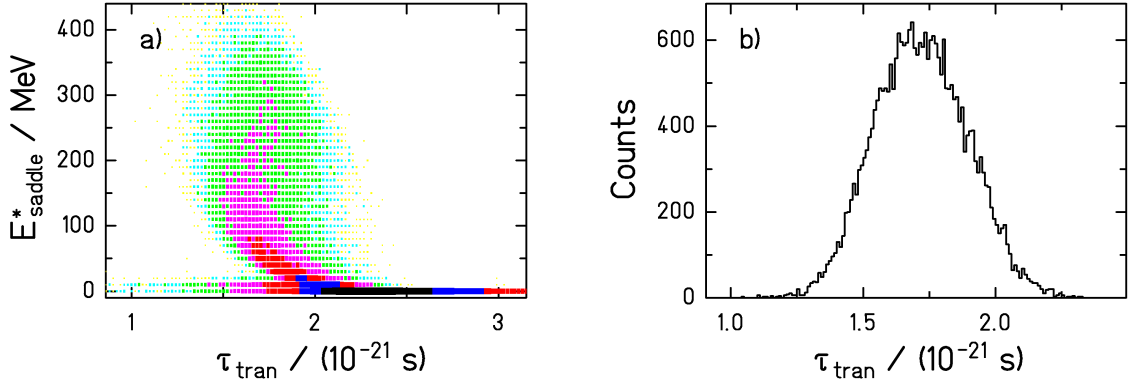

**Figure 15:** a) Excitation energy at fission as a function of the transient time $\tau_{tran}$ calculated according to equation (4) for the reaction $^{238}$U (1 $A$ GeV)+Pb. b) Distribution of transient times corresponding to the fission events with excitation energies at saddle larger than 150 MeV. The calculation has been performed including the break-up stage and using the fission width that follows the analytical solution of the FPE [28, 29].

## 6. Summary

In the present work, a new experimental method suited for the study of dissipation up to the saddle-point deformation has been employed. Fission is induced by peripheral heavy-ion collisions at relativistic energies. The fissioning nuclei produced are characterised by small shape distortions and low angular momenta [33]. Moreover, the high excitation energy induced by the fragmentation of the projectile enables us to be sensitive to the transient time. The specific experimental set-up used allowed for determining different observables sensitive to the strength of dissipation from the ground state up to the saddle point. In particular, the target dependence of the total fission cross sections of $^{238}$U at 1 $A$ GeV as well as the partial fission cross sections and the width of the partial charge distributions of the fission fragments for the reaction $^{238}$U at 1 $A$ GeV on $(CH_2)_n$ were measured. The sensitivity of these observables to dissipation was investigated for the first time.

The experimental observables were compared with calculations obtained with an updated version of the Monte-Carlo code ABRABLA [31, 45] that contains three different approximations for the time-dependent fission width: a step function, an exponential-like in-growth function and a more realistic analytical description based on the solution of the FPE. The transition-state model [2] and several ABRABLA calculations for various values of $\beta$ using the analytical approximation to the solution of the FPE [28], were contrasted with the experimental observables. This analysis clearly demonstrated the influence of dissipation on the way to the fission barrier. None of the new observables could be described by the transition-state model. Although the shape of the target dependence could not be quantitatively reproduced with the present version of the abrasion-ablation model, the minimum of the cross section at $Z_{target} = 6$ could only be explained including dissipation. The best description of the whole set of data was obtained for $\beta = 2 \cdot 10^{21}$ s$^{-1}$. This value corresponds to the critical damping and leads to the lowest possible transient time with a value of $\tau_{tran} \approx (1.7 \pm 0.4) \cdot 10^{-21}$s. This result is in agreement with other work [16, 19, 65], sensitive to the same deformation range. The experimental observables are well reproduced assuming a constant value of $\beta$, independent of temperature and deformation. Note that the ordering parameter $Z_1 + Z_2$ selects nuclei according to their excitation energy, but also according to their fissility, which is connected with the deformation at saddle.

When comparing the model calculations with the new observables, it was found that the exponential-like approximation for the fission-decay width clearly overestimates the width of the charge distributions of the fission fragments. The reason for this discrepancy lies in the fact that such an exponential-like in-growth of the fission width does not imply the strong suppression of fission at very high excitation energies required by the FPE. On the contrary, the step-function approximation and the more realistic description we recently developed showed a quite similar behaviour in very good agreement with the new experimental observables. This indicates that the inhibition of the fission decay width during the initial time laps is needed to account for dissipation effects in a proper way.

## Acknowledgement

We thank D. Boilley and A. V. Ignatyuk for valuable discussions as well as P. Armbruster for careful reading of the manuscript. One of us (C. S.) is grateful for the financing of the stay at GSI, Darmstadt, by the Humboldt foundation. Financial support by the European Community through the HINDAS project under contract FIKW-CT-2000-0031 and by the Spanish MCyT under contract FPA2002-04181-C04-01 is gratefully acknowledged. The work profited from a collaboration meeting on "Fission at finite thermal excitations" in April, 2002, sponsored by the ECT* ("STATE" contract).

## References

[1] H. A. Kramers, Physika VII 4 (1940) 284.
[2] N. Bohr, J. A. Wheeler, Phys. Rev. 56 (1939) 426.
[3] A. Gavron, J. R. Beene, R. L. Ferguson, F. E. Obenshain, F. Plasil, G. R. Young, G.A. Petitt, M. Jääskeläinen, D. G. Sarantites, C. F. Maguire, Phys. Rev. Lett 47 (1981) 1255. Erratum: Phys. Rev. Lett. 48 (1982) 835.
[4] D. Hilscher, E. Holub, U. Jahnke, H. Orf, H. Rossner, Proc. of the 3rd Adriatic Europhysics Conference on the Dynamics of Heavy-Ion Collisions, Hvar, Croatia, Yugoslavia, May 25-30 (1981) 225.
[5] H. Goutte, P. Casoli, J. F. Berger, 5$^{th}$ Seminar on Fission, Point d'Oye (Belgium), Sept. 2003, Proceedings to be published.
[6] H. A. Weidenmüller, Progress in particle and Nuclear Physics, Pergamon, Oxford, Vol. 3 (1980) 49.
[7] H. Risken, The Fokker-Planck Equation, Springer, Berlin (1989) ISBN 0-387-50498.
[8] P. Grangé, L. Jun-Qing, H. A. Weidenmüller, Phys. Rev. C 27 (1983) 2063.
[9] K.-H. Bhatt, P. Grangé, B. Hiller, Phys. Rev. C 33 (1986) 954.

[10] D. Hilscher, I. I. Gontchar, H. Rossner, Yadernaya Fizika 57 (1994) 1255 (Physics of Atomic Nuclei 57 (1994) 1187).
[11] H. van der Ploeg, J. C. S. Bacelar, I. Diószegi, G. van't Hof, A. van der Woude, Phys. Rev. Lett. 75 (1995) 970.
[12] I. Diószegi, N. P. Shaw, A. Bracco, F. Camera, S. Tettoni, M. Mattiuzzi, P. Paul, Phys. Rev. C 63 (2001) 014611.
[13] P. Fröbrich, I. I. Gontchar, N. D. Mavlitov, Nucl. Phys. A 556 (1993) 281.
[14] N. P. Shaw, I. Diószegi, I. Mazumdar, A. Buda, C. R. Morton, J. Velkovska, J. R. Beene, D. W. Stracener, R. L. Varner, M. Thoennessen, P. Paul, Phys. Rev. C 61 (2000) 044612.
[15] A. Saxena, D. Fabris, G. Prete, D. V. Shetty, G. Viesti, B. K. Nayak, D. C. Biswas, R. K. Choudhury, S. S. Kapoor, M. Barbui, E. Fioretto, M. Cinausero, M. Lunardon, S. Moretto, G. Nebbia, S. Pesente, A. M. Samant, A. Brondi, G. La Rana, R. Moro, E. Vardaci, A. Ordine, N. Gelli, F. Lucarelli, Phys. Rev. C 65 (2002) 064601.
[16] K. X. Jing, L. Phair, L. G. Moretto, Th. Rubehn, L. Beaulieu, T. S. Fan, G. J. Wozniak, Phys. Lett. B 518 (2001) 221.
[17] B. Lott, F. Goldenbaum, A. Böhm, W. Bohne, T. von Egidy, P. Figuera, J. Galin, D. Hilscher, U. Jahnke, J. Jastrzebski, M. Morjean, G. Pausch, A. Péghaire, L. Pienkowski, D. Polster, S. Proschitzki, B. Quednau, H. Rossner, S. Schmid, W. Schmid, Phys. Rev. C 63 (2001) 034616.
[18] J. Benlliure, P. Armbruster, M. Bernas, A. Boudard, T. Enqvist, R. Legrain, S. Leray, F. Rejmund, K.-H. Schmidt, C. Stéphan, L. Tassan-Got, C. Volant, Nucl. Phys. A 700 (2002) 469.
[19] S. K. Hui, C. R. Bhuinya, A. K. Ganguly, N. Madhavan, J. J. Das, P. Sugathan, D. O. Kataria, S. Murlithar, Lagy T. Baby, Vandana Tripathi, Akhil Jhingan, A. K. Sinha, P.V. Madhusudhana Rao, N. V. S. V. Prasad, A. M. Vinodkumar, R. Singh, M. Thoennessen, G. Gervais, Phys. Rev. C 62 (2000) 054604, Comment: Phys. Rev. C 64 (2001) 019801, Reply: Phys. Rev. C 64 (2001) 019802.
[20] I. Diószegi, Phys. Rev. C 64 (2001) 019801.
[21] D. Hilscher, H. Rossner, Ann. Phys. Fr. 17 (1992) 471.
[22] Paul, M. Thoennessen, Ann. Rev. Nucl. Part. Sci. 44 (1994) 65.
[23] W. U. Schröder, J. R. Huizenga, "Treatise on Heavy-Ion Science", ed. D. A. Bromley, Vol. 2, Plenum Press New York and London (1984) 115.
[24] P. Hofmann, A. S. Iljinov, Y. S. Kim, M. V. Mebel, H. Daniel, P. David, T. von Egidy, T. Haninger, F. J. Hartmann, J. Jastrzebski, W. Kurzewicz, J. Lieb, H. Machner, H. S. Plendl, G. Riepe, B. Wright, K. Ziock, Phys. Rev. C 49 (1994) 2555.
[25] Y. S. Kim, A. S. Iljinov, M. V. Mebel, P. Hofmann, H. Daniel, T. von Egidy, T. Haninger, F. J. Hartmann, H. Machner, H. W. Plendl, G. Riepe, Phys. Rev. C 54 (1996) 2469.
[26] W. Schmid, T. von Egidy, F. J. Hartmann, J. Hoffmann, S. Schmid, D. Hilscher, D. Polster, H. Rossner, A. S. Iljinov, M. V. Mebel, D. I. Ivanov, V. G. Nedorezov, A.S. Sudov. H. Machner, H. S. Plendl, J. Eades, S. Neumaier, Phys. Rev. C 55 (1997) 2965.
[27] E-M. Eckert, A. Kühmichel, J. Pochodzalla, K. D. Hildenbrand, U. Lynen, W. F. J. Müller, H. J. Rabe, H. Sann, H. Stelzer, W. Trautmann, R. Trockel, R. Wada, A. Cerruti, P. Lhénoret, R. Lucas, C. Mazur, C. Ngô, M. Ribrag, E. Tomasi, A. Demeyer, D. Guinet, Phys. Rev. Lett. 64 (1990) 2483.
[28] B. Jurado, K.-H. Schmidt, J. Benlliure, Phys. Lett. B 533 (2003) 186 (arXiv/nucl-ex/0212020).
[29] B. Jurado, C. Schmitt, K.-H. Schmidt, J. Benlliure, A. R. Junghans, arXiv/nucl-ex-0302003

[30] J. D. Bowman, W. J. Swiatecki, C. E. Tsang, Lawrence Berkeley Laboratory Report LBL-2908 (1973).
[31] J.-J. Gaimard, K.-H. Schmidt, Nucl. Phys. A 531 (1991) 709.
[32] K.-H. Schmidt, T. Brohm, H.-G. Clerc, M. Dornik, M. Fauerbach, H. Geissel, A. Grewe, E. Hanelt, A. Junghans, A. Magel, W. Morawek, G. Münzenberg, F. Nickel, M. Pfützner, C. Scheidenberger, K. Sümmerer, D. Vieira, B. Voss, C. Ziegler, Phys. Lett. B 300 (1993) 313.
[33] M. de Jong, A. V. Ignatyuk, K.-H. Schmidt, Nucl. Phys. A 613 (1997) 435.
[34] F. Goldenbaum, W. Bohne, J. Eades, T. v. Egidy, P. Figuera, H. Fuchs, J. Galin, Ye. S. Golubeva, K. Gulda, D. Hilscher, A. S. Iljinov, U. Jahnke, J. Jastrzebski, W. Kurcewicz, B. Lott, M. Morjean, G. Pausch, Peghaire, L. Pienkowski, D. Polster, S. Proschitski, B. Quednau, H. Rossner, S. Schmid, W. Schmid, P. Ziem, Phys.Rev. Lett. 77 (1996) 1230.
[35] A. Boudard, J. Cugnon, S. Leray, C. Volant, Phys. Rev. C 66 (2002) 044615.
[36] K.-H. Schmidt, M. V. Ricciardi, A. Botvina, T. Enqvist, Nucl. Phys. A 710 (2002) 157 (arXiv/nucl-ex/0207014).
[37] B. Voss, T. Brohm, H.-G. Clerc, A. Grewe, E. Hanelt, A. Heinz, M. de Jong, A. Junghans, W. Morawek, C. Röhl, S. Steinhäuser, C. Ziegler, K.-H. Schmidt, K.-H. Behr, H. Geissel, G. Münzenberg, F. Nickel, C. Scheidenberger, K. Sümmerer, A. Magel, M. Pfützner, Nucl. Instr. Meth. A 364 (1995) 150.
[38] M. Pfützner, H. Geissel, G. Münzenberg, F. Nickel, C. Scheidenberger, K.-H. Schmidt, K. Sümmerer, T. Brohm, B. Voss, H. Bichsel, Nucl. Instr. Meth. B 86 (1994) 213.
[39] K.-H. Schmidt, S. Steinhäuser, C. Böckstiegel, A. Grewe, A. Heinz, A. R. Junghans, J. Benlliure, H.-G. Clerc, M. de Jong, J. Müller, M. Pfützner, B. Voss, Nucl. Phys. A 665 (2000) 221.
[40] S. Steinhäuser, J. Benlliure, C. Böckstiegel, H.-G. Clerc, A. Heinz, A. Grewe, M. de Jong, A. R. Junghans, J. Müller, M. Pfützner, K.-H. Schmidt, Nucl. Phys. A 634 (1998) 89.
[41] F. Rejmund, A. V. Ignatyuk, A. R. Junghans, K.-H. Schmidt, Nucl. Phys. A 678 (2000) 215.
[42] B. Jurado, Ph D. Thesis, University of Santiago de Compostela, June 2002, http://www-w2k.gsi.de/kschmidt/theses.htm.
[43] K. Sümmerer, B. Blank, Phys. Rev. C, 61 (2000) 034607.
[44] J. Cugnon, C. Volant, S. Vuillier, Nucl. Phys. A 620 (1997) 475.
[45] A. R. Junghans, M. de Jong, H.-G. Clerc, A. V. Ignatyuk, G. A Kudyaev, K.-H. Schmidt, Nucl. Phys. A 629 (1998) 635.
[46] T. Enqvist, W. Wlazlo, P. Armbruster, J. Benlliure, M. Bernas, A. Boudard, S. Czajkowski, R. Legrain, S. Leray, B. Mustapha, M. Pravikoff, F. Rejmund, K.-H. Schmidt, C. Stéphan, J. Taïeb, L. Tassan-Got, C. Volant, Nucl. Phys. A 686 (2001) 481.
[47] A.Ya. Rusanov, M. G. Itkis, V. N. Okolovich, Yad. Fiz. 60 (1997) 773 (Phys. At. Nucl. 60 (1997) 683).
[48] S. I. Mulgin, K.-H. Schmidt, A. Grewe, S. V. Zhdanov, Nucl. Phys. A 640 (1998) 375.
[49] D. V. Vanin, G. I. Kosenko, G. D. Adeev, Phys. Rev. C 59 (1999) 2114.
[50] L. F. Oliveira, R. Donangelo, J. O. Rasmussen, Phys. Rev. C 19 (1979) 826.
[51] E. Hanelt, A. Grewe, K.-H. Schmidt, T. Brohm, H.-G. Clerc, M. Dornik, M. Fauerbach, H. Geissel, A. Magel, G. Münzenberg, F. Nickel, M. Pfützner, C. Scheidenberger, M. Steiner, K. Sümmerer, B. Voss, M. Weber, J. Weckenmann, C. Ziegler, Z. Phys. A 346 (1993) 43.
[52] X. Campi, H. Krivine, E. Plagnol, Phys. Rev. C 50 (1994) 2680.
[53] J. Pochodzalla, W. Trautmann, in "Isospin Physics in Heavy-Ion Collisions at Intermediate Energies" Nova Science Publishers, Inc. (2000), arXiv/nucl-ex/0009016.
[54] M. V. Ricciardi, Ph. D. Thesis, University of Santiago de Compostela, Spain, in preparation.

[55] L. G. Moretto, Proc. Of the third IAEA Symposium on the Physics and Chemistry of Fission, Rochester, NY, 13-17 August 1973, IAEA Vienna (1974) , Vol. 1 p. 329.
[56] E. M. Rastopchin, S. I. Mul´gin, Yu. B. Ostapenko, V. V. Pashkevich, M. I. Svirin, G. N. Smirenkin, Yad. Fiz. 53 (1991) 1200 [Sov. J. Nucl. Phys. 53 (1991) 741.
[57] R. Butsch, D. J. Hofman, C. P. Montoya, P. Paul, M. Thoennessen, Phys. Rev. C 44 (1991) 1515.
[58] A. V. Ignatyuk, M. G. Itkis, V. N. Okolovich, G. N. Smirenkin, A. S. Tishin, Yad. Fiz. 21 (1975) 1185 (Sov. J. Nucl. Phys. 21 (1975) 612).
[59] A. V. Karpov, P. N. Nadtochy, E. G. Ryabov, G. D. Adeev, J. Phys. (London) G 29 (2003) 2365
[60] W. D. Myers, W. J. Swiatecki, Ann. Phys. 84 (1974) 186.
[61] A. J. Sierk, Phys. Rev. C 33 (1986) 2039.
[62] Th. Rubehn, R. Bassini, M. Begemann-Blaich, Th. Blaich, A. Ferrero, C. Groß, G. Immé, I. Iori, G. J. Kunde, W. D. Kunze, V. Lindenstruth, U. Lynen, T. Möhlenkamp, L. G. Moretto, W. F. J. Müller, B. Ocker, Pochodzalla, G. Raciti, S. Reito, H. Sann, A. Schüttauf, W. Seidel, V. Serfling, W. Trautmann, A. Trzcinski, G. Verde, A. Wörner, E. Zude, B. Zwieglinski, Phys. Rev. C 53 (1996) 3143.
[63] B. Jurado, C. Schmitt, K.-H. Schmidt, J. Benlliure, A. R. Junghans, arXiv/nucl-ex-0403004
[64] X. Campi, S. Stringari, Z. Phys. A. 309 (1983) 239.
[65] L. G. Moretto, K. X. Jing, R. Gatti, G. J. Wozniak, R. P. Schmitt, Phys. Rev. Lett. 75 (1995) 4186.
[66] B. Jurado, A. Heinz, A. Junghans, K.-H. Schmidt, J. Benlliure, T. Enqvist, F. Rejmund, Proceedings of the XXXIX International Winter meeting on nuclear Physics, January 2001, Bormio, Italy, p. 238. Edited by I. Iori and A. Moroni.